\documentclass[fleqn,twoside]{article}
\usepackage{espcrc2}
\usepackage{epsfig}
\usepackage{graphicx}
\usepackage[figuresright]{rotating}

\newcommand{\AmS}{{\protect\the\textfont2
  A\kern-.1667em\lower.5ex\hbox{M}\kern-.125emS}}

\title{Search for Leptoquarks}

\author{A. Sch\"oning\address[ETH]{Institute of Particle Physics, \\ 
        Eidgen\"ossische  Technische Hochschule Z\"urich, \\ 
        ETH-H\"onggerberg, CH-8093 Z\"urich}}
\begin{document}

\begin{abstract}
This talk 
%(given at the conference ``The legacy of LEP and SLC'', Siena
%Oct. 2001)  
summarizes the status of Leptoquark searches performed at
Tevatron, LEP and HERA. Prospects for Leptoquark searches at future
colliders at LHC and TESLA are given.
\vspace{1pc}
\end{abstract}

% typeset front matter (including abstract)
\maketitle

\section{Introduction}
In many extensions of the Standard Model (SM) new particles are
predicted 
% in the framework of Grand Unification Theories (GUT) 
which
carry both lepton and baryon number - leptoquarks (LQ). LQs might have
masses as low as the electroweak scale. Therefore they have been
directly searched for in high energy collider experiments.
In the BRW model~\cite{ref:brw} LQs are assumed to respect $SU(3) \times
SU(2) \times U(1)$ symmetry. 
%They are grouped in several isospin multiplets. They might exist 
Several isospin multiplets can be constructed
with fermion number $F=\pm 2$ or $F=0$ depending on their
couplings to matter and antimatter, and with spin 0 
(scalar) or spin 1 (vector).

The coupling between LQs and fermions is given by the Yukawa
coupling $\lambda$.
In the most general model couplings of LQs to different fermion
generations can mix and therefore would lead to FCNC and LFV
processes.
In low energy experiments the existence of LQs can be
indirectly tested by interpreting results in the four-fermion
contact interaction model.
Constraints are obtained from the proton lifetime, 
rare decays (e.g. neutrinoless double beta decay), 
atomic parity violation and search for FCNC
processes.
LQs are not allowed to have diquark couplings and have to be chiral
particles, i.e. $\lambda_L \lambda_R \approx 0$.
The ratio of LQ mass over Yukawa coupling $m_{LQ}/\lambda$ has
to be above about 1~TeV in general and above 1-100~TeV for 
flavor changing couplings.
%  \cite{ref:low_energy}.

Recent experiments on the anomalous magnetic moment
of the muon $\Delta a_\mu$ \cite{ref:gmu}\footnote{Recent calculations
of $\Delta a_\mu$ discovered a sign error in the pseudoscalar pole 
contribution, which reduces the deviation \cite{ref:amu_recent}.} and on the atomic parity violation in Cs
\cite{ref:apv} which showed slight deviations from the SM
expectation have been interpreted in the framework of new
LQs by \cite{ref:Barger,ref:Kingman}. These deviations
would be in agreement with the expectation from
LQs of mass about 1 and 1-2~TeV, respectively.

\section{Direct Searches}

In the following the status of experimental searches for first or
second generation LQs at high energy
colliders is summarized. Flavor-changing processes induced by LQs
are not discussed here (for a more general overview on
flavor-changing processes see \cite{ref:talk_fcnc})

\subsection{Tevatron}
\begin{figure}[t]
   \begin{picture}(0,160) 
	\put(15,-20){\epsfig{file=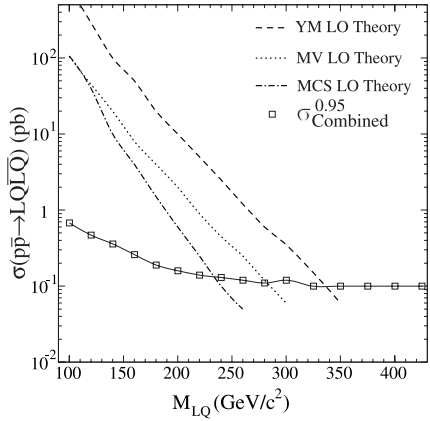,width=0.35\textwidth}}
   \end{picture}
	\caption{The 95\% CL upper limits on cross sections for first
	generation vector LQ pair production as function of
	the LQ mass and comparison
	with LO predictions from different models \protect{\cite{ref:d0_first_gen}}.}
	\label{fig:d0_vector_lq}
\end{figure}

At the Tevatron collider protons and antiprotons are collided at a
center of mass energy of 1.8~TeV.
LQs of all generations would be dominantly
pair-produced independent of the Yukawa
couplings. Limits obtained by a 
search for a first-generation vector LQ performed at the D0
experiment \cite{ref:d0_first_gen} are shown in
Fig.~\ref{fig:d0_vector_lq}. 
LQs with
a branching ratio $\textrm{BR}(LQ \rightarrow eq,\nu q)=1/2$ were
considered in the BRW model by combining the searches for the 
resulting $eejj$, $e\nu
jj$ and $\nu \nu jj$ channels. The results were compared with
different assumptions on the LQ-LQ gauge couplings: Yang
Mills, minimal vector, minimal cross section couplings.
Depending on the model 95\% CL limits have been set in the range
230-340~GeV. In a recent publication similar limits were
obtained from a search for the $\nu \nu jj$ channel alone assuming 
$\textrm{BR}(LQ \rightarrow \nu q)=1$ \cite{ref:d0_recent}.

\begin{figure}[t]
   \begin{picture}(0,160) 
	\put(15,-25){\epsfig{file=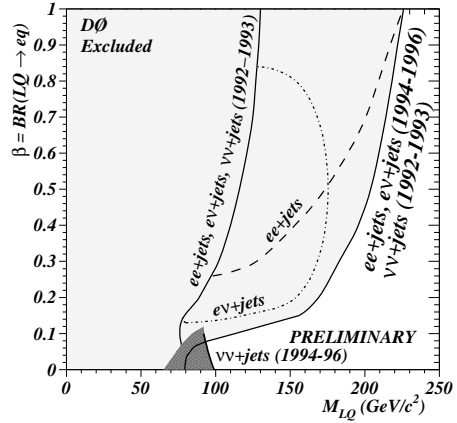,width=0.4\textwidth}}
   \end{picture}
	\caption{The 95\% CL lower limit on the mass of
	first-generation scalar LQs as a function of the
	Branching Ratio $\beta$ for the individual $eejj$, $e\nu jj$,
	and $\nu \nu jj$ channels, and for the combined analysis
	\protect{\cite{ref:d0_first_gen}}.} 
	\label{fig:d0_scalar_lq}
\end{figure}

The results of a search for pair production of a first-generation
scalar LQ is summarized in Fig.~\ref{fig:d0_scalar_lq}.
LQs are here interpreted in a more general model with
arbitrary branching ratios into $eq$ and $\nu q$.
For branching ratios with a dominant decay into $eejj$ masses below
about 225~GeV,  for branching ratios with a dominant decay into $\nu \nu jj$ masses below
about 95-100~GeV are excluded.

Searches for second generation vector LQs yielded due to the cleaner
signature of muons in the final state higher mass limits. 
Depending on the branching ratio into muon and neutrino 
masses above about 270-320~GeV were excluded.

\subsection{LEP}
In $e^+e^-$ collisions LQs can be tested in both
pair production and single production. In pair production the
sensitivity is limited to half the center of mass energy $m_{LQ} <
\sqrt{s}/2$. Therefore mass exclusion limits obtained at LEP II \cite{ref:lep_pair}
of the order of 100~GeV are not competitive with those obtained by
Tevatron.
In single production LQs can be created up to the full center
of mass energy. 
LQs are mainly photo-produced and the cross section depends
quadratically on the Yukawa coupling $\lambda$. 
Searches have been performed \cite{ref:lep_single} for several
types of LQs in different decay channels. Limits on $\lambda$
of the order 0.1-1 were obtained in the mass range 100-189~GeV for
scalar and vector LQs.

\subsection{HERA}
\begin{figure}[t]
   \begin{picture}(0,160) 
	\put(15,-20){\epsfig{file=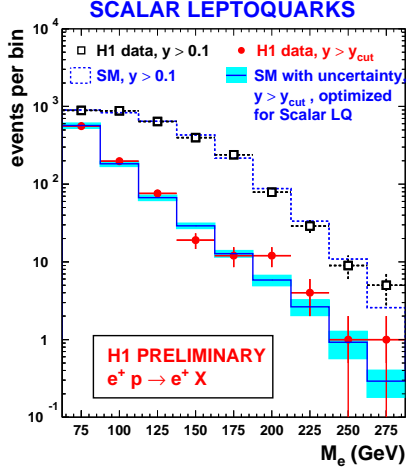,width=0.38\textwidth}}
   \end{picture}
	\caption{Distribution of {\em positron-jet} invariant masses
	measured by the H1 experiment. The open squares show all
	data, the full points show the data after applying an angular
	optimization cut for {\em scalar LQs}. The lines indicate 
	the expectation from the SM}
	\label{fig:h1_scalar_lq_mass}
\end{figure}

At the electron(positron)-proton collider HERA LQs can be
resonantly produced in $e^\pm q$ interactions up to the center of
mass energy of $\sqrt{s}=318~\textrm{GeV}$ for not too small values of
$\lambda$.
The main background is due to genuine neutral current
(NC) and charged current (CC) deep inelastic scattering processes.
Different types of LQs can in principle be tested by
switching the lepton beam charge and by studying the decay
distributions of LQs. The latter has been exploited to
discriminate against SM background and to enhance sensitivity.
The distribution of reconstructed invariant masses as
measured by the H1 experiment is shown in
Fig.~\ref{fig:h1_scalar_lq_mass}.
Optimization cuts have been applied to enhance the sensitivity for
scalar LQs.

An excess found earlier at masses of about 200~GeV in the data taken from
1994-1997 by the H1 and ZEUS collaborations \cite{ref:hera_lq_excess}
was not confirmed by the most recent data. The
slight excess at 200~GeV in Fig.~\ref{fig:h1_scalar_lq_mass}
 is still a remaining effect of the early HERA excess.

\begin{figure}[t]
   \begin{picture}(0,160) 
	\put(15,-20){\epsfig{file=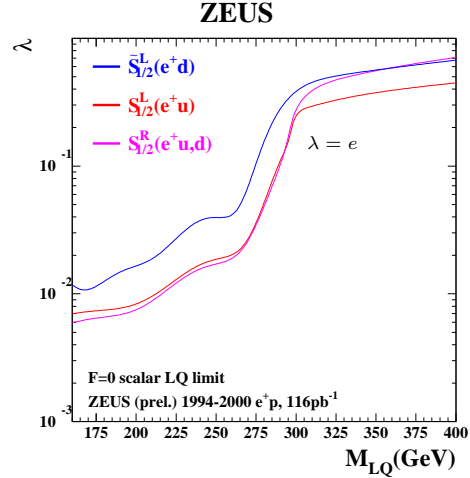,width=0.41\textwidth}}
	\put(135,104){\scriptsize $\lambda=e$}
   \end{picture}
	\caption{95\% CL exclusion limits of scalar LQs as function
	of the mass calculated in the BRW model. Yukawa couplings above the curves are excluded.}
	\label{fig:zeus_lq_scalar}
\end{figure}
Similar results obtained by the ZEUS collaboration from data taken 
in $e^+p$ scattering have been interpreted in the BRW model.
Limits are shown in Fig.~\ref{fig:zeus_lq_scalar} 
considering different types of $F=0$ scalar LQs. 
For LQ masses above 200~GeV Yukawa
couplings of the order 0.01 can be excluded almost up to the center of
mass energy. For Yukawa couplings of electromagnetic strength
$\lambda=e\approx 0.3$ scalar LQs up to about 290-300~GeV are excluded.

\begin{figure}[t]
   \begin{picture}(0,120) 
	\put(-5,-30){\epsfig{file=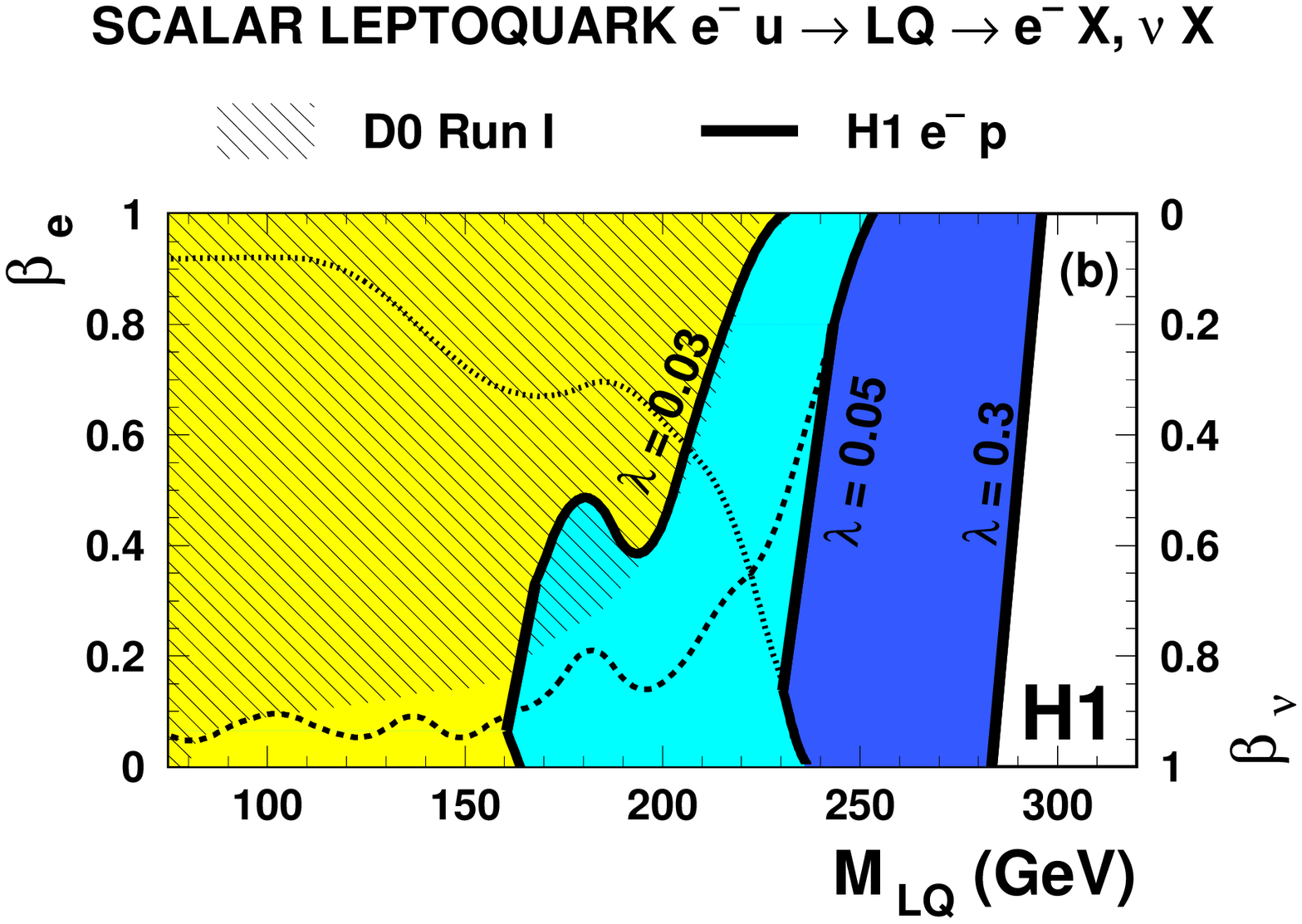,width=0.5\textwidth}}
   \end{picture}
	\caption{95\% CL mass exclusion limits of a $F=2$ scalar LQ as function
	of the branching ratio $\textrm{BR}(LQ \rightarrow e q)$. The
	limits are shown for three different example Yukawa couplings.}
	\label{fig:h1_lq_eu_limit.eps}
\end{figure}
The exclusion limit obtained by the H1 experiment in a more general
model with an arbitrary branching ratio into electron and neutrino of a
$F=2$ scalar LQ is shown in Fig.~\ref{fig:h1_lq_eu_limit.eps}.
Due to the combination of the NC and CC channels the 
limits, which are shown for three different assumptions on $\lambda$,
are almost independent of the branching ratio.

\begin{figure}[t]
   \begin{picture}(0,170) 
	\put(0,-30){\epsfig{file=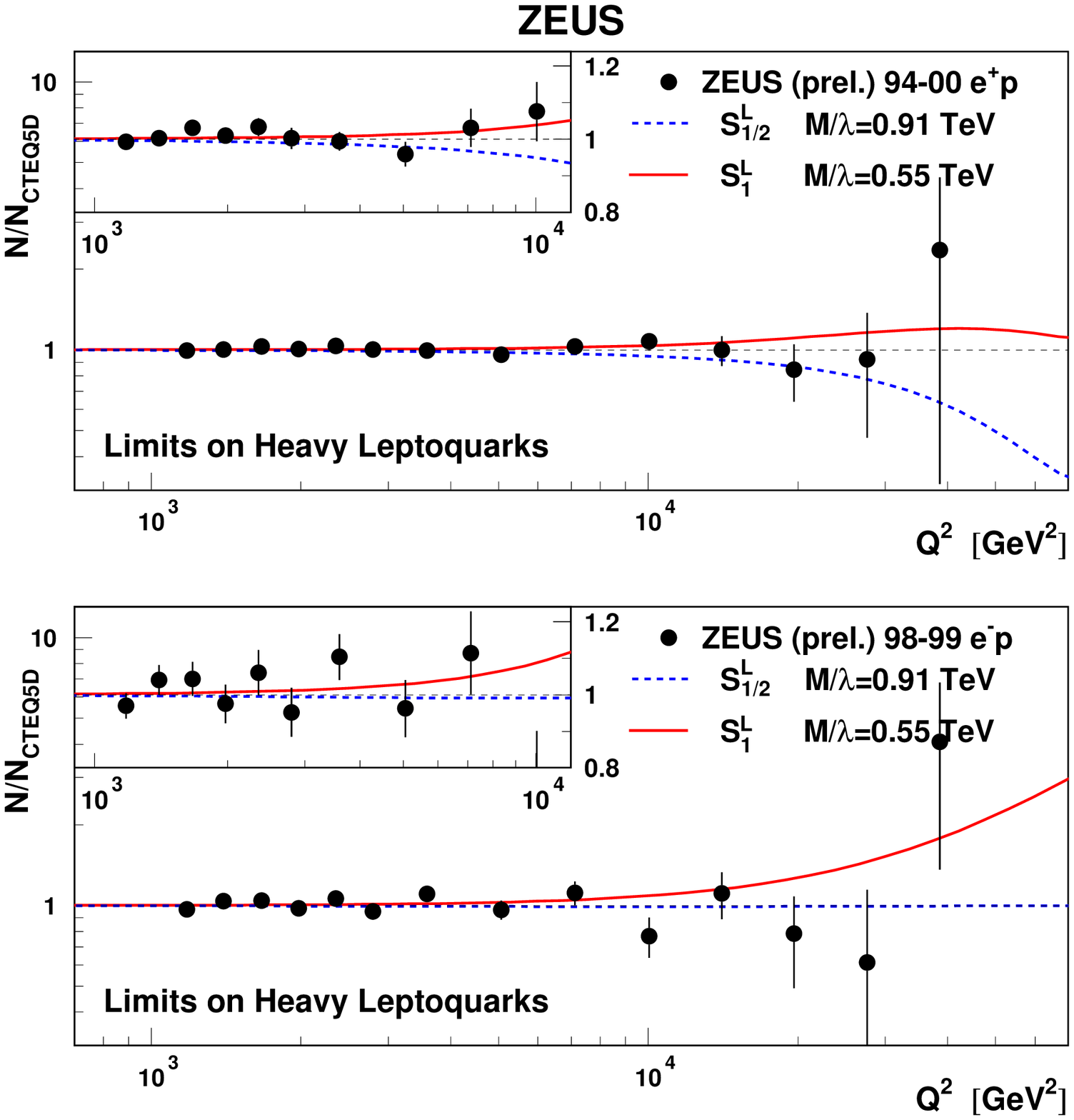,width=0.45\textwidth}}
   \end{picture}
	\caption{95\% CL Limits on LQs. The full lines show limits of
	two specific LQs obtained by a simultaneous fit to $e^+p$ and
	$e^-p$ scattering data as function of $Q^2$ \protect{\cite{ref:zeus_ci}}.}
	\label{fig:zeus_ci_lqfit}
\end{figure}

\section{Contact Interaction}
In the framework of four-fermion contact interactions additional
LQ couplings can be studied at different colliders by testing
 the $\ell\ell qq$ vertex at high scales
(e.g. $Q^2$). 
By adding to the SM Lagrangian an
extra term with positive or negative interference $\varepsilon=\pm1$ for the contact interaction
deviations from the SM can be interpreted as function of $M/\lambda$~\footnote{
often the convention $\Lambda/g$ is used} with $M$ being the mass scale.

The H1 and ZEUS experiments at HERA have fitted various LQ hypotheses to the NC
data taken at HERA~I. Lower limits at 95\% CL on $M/\lambda$ for two
specific LQs are shown in Fig.~\ref{fig:zeus_ci_lqfit}. Typical limits
for different types of LQs obtained at HERA are in the range
$0.5-1.3$~TeV \cite{ref:zeus_ci,ref:h1_ci}.

\begin{figure}[t]
   \begin{picture}(0,220) 
	\put(-5,-40){\epsfig{file=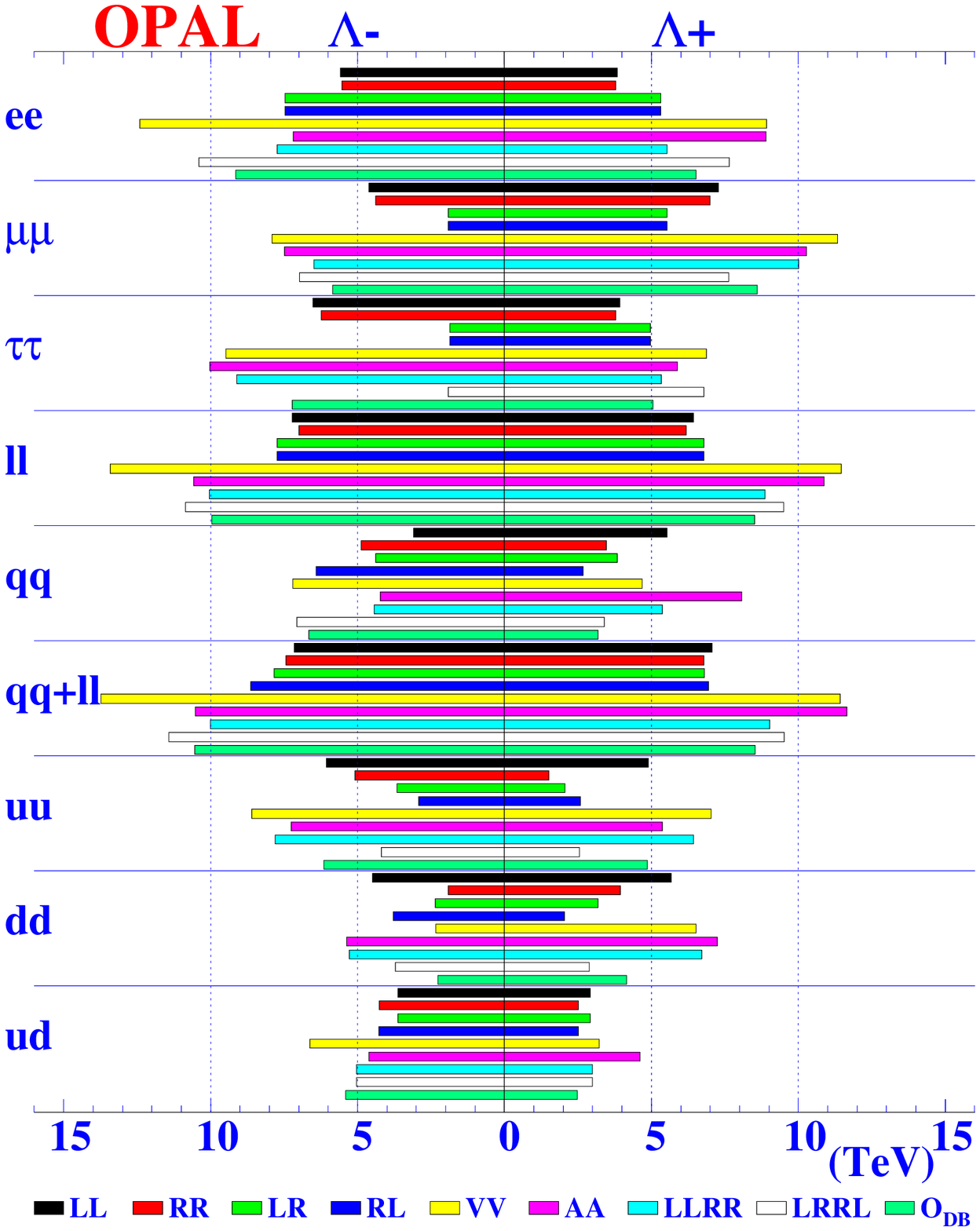,width=0.5\textwidth}}
   \end{picture}
	\caption{95\% limits on an effective contact interaction with
	a mass scale $\Lambda$ for different final state topologies  \protect{\cite{ref:opal_ci}}.}
	\label{fig:opal_ci}
\end{figure}

Similarly, different kinds of  contact interactions were fitted to the
two-fermion final state cross sections at LEP~II. The resulting 95\%~CL 
limits as obtained by the OPAL collaboration~\cite{ref:opal_ci}
for different chiral couplings are shown in Fig.~\ref{fig:opal_ci}.
The limits are given here as lower limit on $\Lambda/g$ with
$g=\sqrt{4\pi}$ for both constructive and destructive interference with
the SM processes. The limits on $q\bar{q}$ final states can also be
interpreted in LQ models resulting in mass limits similar
to those obtained at HERA.

\section{Summary}
\begin{figure}[t]
   \begin{picture}(0,170) 
	\put(0,-30){\epsfig{file=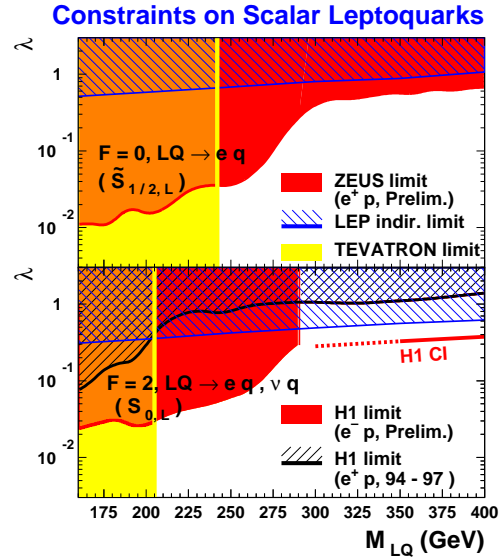,width=0.45\textwidth}}
   \end{picture}
	\caption{95\%~Limits from LQ searches obtained from collider
	experiments at Tevatron, HERA and LEP on $\lambda$ as
	function of the LQ mass. }
	\label{fig:exp_summary}
\end{figure}

The current status of first-generation LQ searches is
summarized in Fig.~\ref{fig:exp_summary}  for two specific
scalar LQ examples. The plot shows excluded regions as function of the
LQ mass and its coupling as obtained by direct and indirect searches performed at
Tevatron, HERA and LEP. 
Tevatron searches are independent  of the Yukawa coupling while HERA
searches have a higher mass reach for not too small Yukawa couplings. 
LQs at very  high masses above 300~GeV are only constrained 
from indirect searches performed at HERA and LEP.

\section{Outlook}
Both HERA and Tevatron were upgraded in order to produce more
luminosity. 
In the next years before LHC and TESLA come to
operation an exciting competition between HERA and Tevatron
in scalar LQ searches up to 300~GeV is expected. 
In case of a discovery HERA has the possibility
by using different beam charges and by exploiting the newly available
longitudinally polarized $e^\pm$ beam to determine the LQ type.
For not too small Yukawa couplings LQs may be tested with higher sensitivity in
indirect searches at the TeV scale. This region will be directly
probed by LHC were resonances up to about 1.5~TeV can be found \cite{ref:lhc}.
The successor of LEP, TESLA, will be able to probe LQs up to about
350~GeV. In case of a discovery TESLA will have the possibility to
determine precisely the LQ type, its mass and its couplings \cite{ref:tesla}.

\end{document}